\documentclass[fleqn,usenatbib]{mnras}

\usepackage[T1]{fontenc}
\usepackage{ae,aecompl}

\usepackage{graphicx}
\usepackage{amsmath}
\usepackage{amssymb}
\usepackage{nicefrac}
\usepackage{multirow}
\usepackage{threeparttable}
\usepackage{footmisc}
\usepackage{txfonts}

\newcommand{\sub}[1]{{\rm #1}}
\newcommand{\solarmass}{\, \text{M}_\odot}
\newcommand{\solarmassperyear}{\text{M}_\odot/\text{yr}}

\defcitealias{2016MNRAS.462..250M}{M16}

\title[Predictions of IRX and cosmic SFRD]{Dark-ages Reionization and Galaxy Formation Simulation - XIX: Predictions of infrared excess and cosmic star formation rate density from UV observations}

\author[Qiu et al.]{Yisheng Qiu$^{1,2}$\thanks{E-mail: yishengq@student.unimelb.edu.au},
                    Simon J. Mutch$^{1,2}$,
                    Elisabete da Cunha$^{2,3}$,
                    Gregory B. Poole$^{4}$,
                    \newauthor
                    J. Stuart B. Wyithe$^{1,2}$\thanks{E-mail: swyithe@unimelb.edu.au}
	\\$^{1}$School of Physics, University of Melbourne, Parkville, VIC 3010, Australia
	\\$^{2}$ARC Centre of Excellence for All Sky Astrophysics in 3 Dimensions (ASTRO 3D)
	\\$^{3}$Research School of Astronomy and Astrophysics, The Australian National University, Canberra ACT 2611, Australia
	\\$^{4}$Centre for Astrophysics and Supercomputing, Swinburne University of Technology, PO Box 218, Hawthorn VIC 3122, Australia}
\begin{document}

\pagerange{\pageref{firstpage}--\pageref{lastpage}} \pubyear{2017}
\maketitle
\label{firstpage}

\begin{abstract}
We present a new analysis of high-redshift UV observations using a semi-analytic galaxy formation model, and provide self-consistent predictions of the infrared excess (IRX) - $\beta$ relations and cosmic star formation rate density. We combine the Charlot \& Fall dust attenuation model with the \textsc{meraxes} semi-analytic model, and explore three different parametrisations for the dust optical depths, linked to star formation rate, dust-to-gas ratio and gas column density respectively. A Bayesian approach is employed to statistically calibrate model free parameters including star formation efficiency, mass loading factor, dust optical depths and reddening slope directly against UV luminosity functions and colour-magnitude relations at $z \sim 4 - 7$. The best-fit models show excellent agreement with the observations. We calculate IRX using energy balance arguments, and find that the large intrinsic scatter in the IRX - $\beta$ plane correlates with specific star formation rate. Additionally, the difference among the three dust models suggests at least a factor of two systematic uncertainty in the dust-corrected star formation rate when using the Meurer IRX - $\beta$ relation at $z \gtrsim 4$. 
\end{abstract}


\section{Introduction}
One fundamental question in astronomy is to understand the buildup of stars and galaxies from baryonic matter in the early Universe. During this epoch, observations focus mainly on rest-frame UV properties due to cosmic redshift. These include measurements of UV luminosity functions (LFs) \citep{2010A&A...523A..74V,2015ApJ...803...34B,2017ApJ...835..113L,2018arXiv180707580B,2018PASJ...70S..10O}, and UV continuum slope to UV magnitude relations \citep{2012ApJ...756..164F,2014ApJ...793..115B,2014MNRAS.440.3714R}, which are also known as the colour-magnitude relations (CMRs). The UV luminosity is a tracer of star formation since most UV photons are emitted by young stars. However, star formation can be heavily obscured by the interstellar dust. One commonly adopted approach to perform dust corrections at high redshifts is to infer the infrared excess (IRX) from the observed UV slopes using a relation calibrated by \cite{1999ApJ...521...64M} \citep[e.g.][]{2015ApJ...803...34B,2015ApJ...813...21M,2016MNRAS.462..235L}. However, the \cite{1999ApJ...521...64M} relation is calibrated against local starburst galaxies, and observations of far infrared data is rather challenging at high redshifts. Recent observations at $z \gtrsim 3$ show large scatter in the IRX - $\beta$ relation \citep{2015Natur.522..455C,2016A&A...587A.122A,2016ApJ...833...72B,2017ApJ...845...41B,2017MNRAS.472..483F,2018MNRAS.479.4355K}. For instance, the observed IRX by \cite{2016ApJ...833...72B} is much lower than the \cite{1999ApJ...521...64M} relation, while \cite{2018MNRAS.479.4355K} suggest that the IRX - $\beta$ relation does not evolve with redshift. These observations motivate investigation of the IRX - $\beta$ at high redshifts from theoretical models.
\par
Theoretical studies of dust extinction require intrinsic galaxy properties as input, and one approach is to postprocess the output of a hydrodynamical simulation. This method has been implemented in \cite{2017ApJ...840...15S} and \cite{2018MNRAS.474.1718N} to investigate the origin of the IRX - $\beta$ relation. At $z \gtrsim 5$, the IRX - $\beta$ relation has been studied by \cite{2016MNRAS.462.3130M}, \cite{2017MNRAS.470.3006C} and \cite{2019MNRAS.487.1844M}. However, their results suggest different extinction curves. \cite{2017MNRAS.470.3006C} pointed out that the reason for the disagreement could be due to systematics associated with different simulations.
\par
Semi-analytic models are another popular approach for studying galaxy formation \citep[e.g.][]{2011MNRAS.413..101G,2015MNRAS.453.4337S,2016ApJS..222...22C,2016MNRAS.462.3854L,2018MNRAS.479....2C,2018MNRAS.481.3573L,2019arXiv190101906C}, and their results can also be used as input to dust models. Semi-analytic models solve a system of differential equations that govern the mass accretion and transition of several key baryonic components of galaxies such as gas and stellar mass. The construction of these models is relatively simple, and hence they are computationally efficient. These models also introduce several free parameters to describe the unknown efficiency or strength of certain physics processes. These parameters bring flexibility, and allow the exploration of different galaxy formation scenarios, which is very useful for identifying which galaxy processes regulate certain observations.
\par
The Dark-ages Reionization And Galaxy Observables from Numerical Simulations (DRAGONS) project\footnote[1]{\label{dragons}http://dragons.ph.unimelb.edu.au/} introduces the \textsc{meraxes} semi-analytic model \citep{2016MNRAS.462..250M}, which is coupled with the high cadence \textit{Tiamat} N-body simulation \citep{2016MNRAS.459.3025P}. The model concentrates on studying galaxy formation at high redshifts. This work utilises \textsc{meraxes} to predict intrinsic galaxy properties, and combines it with a simple and flexible dust attenuation model. The dust optical depths are calculated empirically using relevant galaxy properties. By taking full advantage of the fast computational speed of both the galaxy formation and dust models, we carry out a Bayesian analysis on all the model free parameters, and use UV LFs and CMRs as constraints, which are the most fundamental observables at high redshift. This approach allows these observations to put direct constraints on both galaxy formation and dust parameters, and provides self-consistent predictions of the IRX and star formation rate (SFR).
\par
We organise the paper as follows. Section \ref{SAM} provides an overview of our \textsc{meraxes} galaxy formation model, and introduces several updates on the model for this work. Section \ref{SED} describes the dust models that are integrated into \textsc{meraxes} and the computation of galaxy spectral energy distributions (SEDs). The description of our calibration method can found in Section \ref{calib}, and the results are discussed in Section \ref{res}. We demonstrate the predicted IRX - $\beta$ relations and cosmic star formation rate density (SFRD) in Section \ref{IRX} and Section \ref{CSFD} respectively. Finally, this work is summarised in Section \ref{sum}. Throughout the paper, we adopt a flat $\Lambda$CDM cosmology, with $(h, \Omega_\text{m}, \Omega_\text{b}, \Omega_\Lambda, \sigma_8, n_\text{s}) = (0.678, 0.308, 0.0484, 0.692, 0.815, 0.968)$ \citep{2016A&A...594A..13P}. Magnitudes are in the AB system \citep{1983ApJ...266..713O}.

\section{Galaxy formation model} \label{SAM}
\subsection{Overview}
The \textsc{meraxes} semi-analytic model\textsuperscript{\ref{dragons}} is the backbone of the present work. It extends the models of \cite{2006MNRAS.365...11C} and \cite{2011MNRAS.413..101G} to high redshifts, and is modified to run on high cadence halo merge trees with a delayed supernova feedback scheme. It also implements gas infall, radiative cooling, star formation, supernova feedback, metal enrichment, and reionisation feedback. A detailed description of the model can be found in \cite{2016MNRAS.462..250M}, hereafter \citetalias{2016MNRAS.462..250M}. The active galactic nuclei (AGN) feedback of the model is later introduced by \cite{2017MNRAS.472.2009Q}. This work also applies several updates to the model, aiming to improve the predicted gas phase metallicity, which is an input of galaxy SEDs. These will be introduced in Section \ref{mod}.
\par
We utilise the halo merger trees of the \textit{Tiamat} N-body simulation \citep{2016MNRAS.459.3025P,2017MNRAS.472.3659P} as input to our semi-analytic model. The simulation contains $2160^3$ particles in a $(67.8 h^{-1})^3 \, \text{Mpc}^3$ box, with mass resolution $m_\sub{p} = 2.64 \times 10^6 h^{-1} \, \solarmass$. Halos and friends-of-friends groups are identified using \textsc{subfind} \citep{2001MNRAS.328..726S}. The timestep of the simulation is $11.1 \, \text{Myr}$ between $z = 35$ and $z = 5$ and is evenly distributed in dynamical time between $z = 5$ and $z = 1.8$. The high cadence of the simulation is critical to this study since UV magnitudes are sensitive to starbursts in the recent 100 Myr.
\par
Since this work requires evaluating the model many times, and does not focus on ionising structures, we adopt homogeneous reionisation feedback \citep{2000ApJ...542..535G} instead of using \textsc{21cmfast} \citep{2007ApJ...669..663M}. Both approaches are described in \citetalias{2016MNRAS.462..250M} and found to have almost the same predictions on global galaxy properties such as the stellar mass function up to $z \simeq 5$. However, the homogeneous prescription is more computationally efficient.

\subsubsection{Star Formation} \label{SF}
The star formation model in \citetalias{2016MNRAS.462..250M} should be mentioned here, since the free parameters in the model will be calibrated statistically in this work. Our model assumes that gas undergoes shock heating and forms a quasi-static hot halo when it is accreted by the host dark matter halo. The gas can cool and form a cold disk in the central region, which then becomes fuel for star formation. The gas in the hot halo and the cold disk is labelled as hot and cold gas respectively. Following the disk stability argument of \cite{1996MNRAS.281..475K}, our model assumes that gas can only form stars when its mass is greater than the critical mass
\begin{equation} \label{m_crit}
    m_\text{crit} = \Sigma_\text{SF} \left( \frac{V_\text{max}}{100 \text{km/s}} \right) \left( \frac{r_\text{disk}}{10 \text{kpc}}  \right) \times 10^{10} \solarmass,
\end{equation}
where $V_\text{max}$ is the maximum circular velocity of the host halo. The disk scale radius $r_\text{disk}$ is defined by
\begin{equation} \label{r_disk}
    r_\text{disk} = 3 R_\text{vir} \frac{\lambda}{\sqrt{2}},
\end{equation} 
where $R_\text{vir}$ is the virial radius of the host halo, and $\lambda$ is the spin parameter defined by \cite{2001ApJ...555..240B}. Then, the mass of new formed stars can be calculated from
\begin{equation} \label{alpha_SF}
    \Delta m_\text{star} = \alpha_\text{SF} \frac{m_\text{gas} - m_\text{crit}}{t_\text{dyn,disk}} \Delta t,
\end{equation}
where $t_\text{dyn, disk} = r_\text{disk}/V_\text{max}$ is the dynamical time of the disk, $m_\text{gas}$ is the mass of cold gas and $\Delta t$ is the timestep. In the model, the normalisation of the critical mass $\Sigma_\text{SF}$ and the star formation efficiency $\alpha_\text{SF}$ are the two free parameters. Their preferred values will be discussed in Section \ref{res}.

\subsection{Updates to Meraxes} \label{mod}
\subsubsection{Supernova Feedback} \label{feedback}
We update the supernova feedback model with a different treatment of supernova energy, and a different parametrisation of mass loading factor and energy coupling efficiency. Our original supernova model in \citetalias{2016MNRAS.462..250M} is a modified version of \cite{2011MNRAS.413..101G}, taking into account the high cadence of our halo merger trees. As mentioned in Section \ref{SF}, our model galaxies have hot and cold gas components, and the effect of supernova feedback is to transfer the gas in the cold disk to the hot halo. The amount of mass that is reheated by supernova can be calculated by
\begin{equation} \label{m_reheat}
  \Delta m_\text{reheat} = \begin{cases}
    \eta \Delta m_\text{new}, & \Delta E_\text{SN} \geq \Delta E_\text{hot} \\
    \frac{\Delta E_\text{SN}}{\nicefrac{1}{2} V_\text{vir}^2}, & \Delta E_\text{SN} < \Delta E_\text{hot}
  \end{cases},
\end{equation}
with
\begin{equation}
    \Delta E_\text{hot} = \frac{1}{2} \eta \Delta m_\text{new} V_\text{vir}^2,
\end{equation}
where $\eta$ is the mass loading factor, $\Delta m_\text{new}$ is the mass of new formed stars, $\Delta E_\text{SN}$ is the supernova energy that is injected into the interstellar medium (ISM), and $V_\text{vir}$ is the virial velocity of the friends-of-friends group. If the amount of reheated mass is $\Delta m$, the energy increase of the hot halo is $\Delta E = \frac{1}{2} \Delta m V_\text{vir}^2$ after virialisation. This model first estimates the reheated mass by the mass loading factor argument, and reduces the mass if the energy injected by supernova is smaller than the underlying energy increase of the hot halo. Moreover, if $\Delta E_\text{SN} \geq \Delta E_\text{hot}$, materials can be further ejected from the hot halo. The amount of ejected mass is given by
\begin{equation}
    \Delta m_\text{eject} = \frac{\Delta E_\text{SN} - \Delta E_\text{hot}}{\nicefrac{1}{2} V_\text{vir}^2}.
\end{equation}
The ejected mass is subtracted from the hot gas and put into a separated component. 
\par
The injected supernova energy $\Delta E_\text{SN}$ plays a important role in the model described above. This quantity is given by
\begin{equation} \label{E_SN}
    \Delta E_\text{SN} = \epsilon \times \int^{t + \Delta t}_t dt' \int^\infty_0 d\tau \, \frac{d\varepsilon}{d\tau} \psi_\text{}(t' - \tau),
\end{equation}
where $\epsilon$ is the energy coupling efficiency, $t$ is the simulation time, $\Delta t$ is the timestep, $(d\varepsilon / d\tau)d\tau$ is the energy released by type-II supernova from stars with age between $\tau$ to $\tau + d\tau$ per unit mass of stellar population, and $\psi_\text{}(t)$ is the star formation rate as a function of the simulation time. The second term on the right hand side of Equation (\ref{E_SN}) is the total energy released by type-II supernova during a snapshot. \citetalias{2016MNRAS.462..250M} uses an analytic fit of star lifetime and an initial mass function to estimate $d\varepsilon / d\tau$, while in this work, we generate $d\varepsilon / d\tau$ using \textsc{starburst99} \citep{1999ApJS..123....3L,2005ApJ...621..695V,2010ApJS..189..309L,2014ApJS..212...14L} with metallicity dependence, assuming a \cite{2002Sci...295...82K} initial mass function (IMF). This treatment provides more reasonable and self-consistent estimates of the supernova energy, and can be generalised to other stellar evolutionary libraries \citep[e.g][]{2017AJ....153...85S,2018ApJS..237...42R}. A similar approach has already been applied in the \textsc{fire} hydrodynamic simulations \citep{2014MNRAS.445..581H}.
\par
To evaluate the integral in Equation (\ref{E_SN}), we adopt the same method as \citetalias{2016MNRAS.462..250M}. \textsc{meraxes} tracks the mass of new formed stars and their metals in four previous snapshots and assumes that they are formed by a single burst in the middle of each corresponding snapshot. Stars formed in earlier snapshots have ages greater than 55 Myr, and typically do not end with a type-II supernova. To tackle the metallicity dependence, we interpolate the table of $d\varepsilon / d\tau$ from \textsc{starburst99} to a grid in a range of $Z = 0.001 \sim 0.040$ with resolution $\Delta Z = 0.001$, and apply nearest interpolation based on the grid for each starburst.
\par
Since supernova energy is released by stars formed in current and several previous snapshots, $\Delta m_\text{new}$ in Equation (\ref{m_reheat}) should have contributions from these stars. This quantity is computed by
\begin{equation} \label{m_new}
    \Delta m_\text{new} = \frac{\int^{t + \Delta t}_t dt' \int^\infty_0 d\tau \, \frac{d\varepsilon}{d\tau} \psi(t' - \tau)}{\int^\infty_0 d\tau \frac{d\varepsilon}{d\tau}}
\end{equation}
In other words, we use the average star formation history weighted by the supernova energy to calculate $\Delta m_\text{new}$. If we assume constant canonical energy for every type-II supernova explosion, the above equation is equivalent to the number-weighted expression given by Equation (16) in \citetalias{2016MNRAS.462..250M}.
\par
The remaining parameters in the supernova feedback model are the mass loading factor $\eta$ and energy coupling efficiency $\epsilon$. In this work, we adopt different parametrisations from \citetalias{2016MNRAS.462..250M}. They are given by
\begin{equation} \label{eta}
    \eta = \begin{cases}
    \eta_0 \left ( \frac{1 + z}{4}\right)^{\alpha_\text{reheat}} \left ( \frac{V_\text{max}}{60 \text{km/s}}\right)^{-1}, & V_\text{max} \geq 60 \text{km/s} \\
    \eta_0 \left ( \frac{1 + z}{4}\right)^{\alpha_\text{reheat}} \left ( \frac{V_\text{max}}{60 \text{km/s}}\right)^{-3.2}, & V_\text{max} < 60 \text{km/s}\\
  \end{cases},
\end{equation}
\begin{equation} \label{epsilon}
    \epsilon = \begin{cases}
    \epsilon_0 \left ( \frac{1 + z}{4}\right)^{\alpha_\text{eject}} \left ( \frac{V_\text{max}}{60 \text{km/s}}\right)^{-1}, & V_\text{max} \geq 60 \text{km/s} \\
    \epsilon_0 \left ( \frac{1 + z}{4}\right)^{\alpha_\text{eject}} \left ( \frac{V_\text{max}}{60 \text{km/s}}\right)^{-3.2}, & V_\text{max} < 60 \text{km/s}\\
  \end{cases},
\end{equation}
where $V_\text{max}$ is the maximum circular velocity. We force the maximum of $\epsilon$ to be unity due to energy conservation. \cite{2015MNRAS.454.2691M} originally obtained a broken power law for the mass loading factor. Their study is based on model galaxies in the \textsc{fire} simulations \citep{2014MNRAS.445..581H}. This form is subsequently implemented in several semi-analytic models \citep{2016MNRAS.461.1760H,2018MNRAS.479....2C,2018MNRAS.481.3573L}. The implementation of this form in the present work is primarily motivated by its impact on the metallicity, which is an input of galaxy SEDs. \citet{2016MNRAS.461.1760H} tested eight different supernova feedback schemes in their semi-analytic model, and found that only explicit redshift-dependent models can lead to evolution of the mass metallicity relation. \cite{2018MNRAS.481..954C} demonstrated that a steeper slope of the redshift dependence can result in stronger evolution of the mass metallicity relation using the semi-analytic model of \cite{2018MNRAS.479....2C}. In this work, we set $\alpha_\text{reheat} = 2$ according to the optimisation result in \cite{2018MNRAS.479....2C}, assume no redshift dependence on the energy coupling efficiency (i.e. $\alpha_\text{eject} = 0$) and leave $\eta_0$ and $\epsilon_0$ as free parameters.
\subsubsection{Mass recycling and metal enrichment}
We also apply \textsc{starburst99} to the mass recycling and metal enrichment. The mass of materials that are produced by type-II supernova and released into the ISM can be obtained by
\begin{equation} \label{m_recycle}
        \Delta m_\text{recycle} = \int^{t + \Delta t}_t dt' \int^\infty_0 d\tau \, \frac{dy}{d\tau} \psi(t' - \tau),
\end{equation}
where $(dy/d\tau)d\tau$ is the mass produced by type-II supernova from stars with age $\tau$ to $\tau + d\tau$ per unit mass of stellar population, i.e. the yield. This quantity depends on the IMF and varies with different elements.
We generate the table of $dy/d\tau$ using \textsc{starburst99}, including metallicity dependence and assuming a \cite{2002Sci...295...82K} IMF. In the present work, we only consider two cases, i.e. the yield of all elements and the yield of all metal elements. The former gives the amount of recycled mass, while the latter introduces metal enrichment. The evaluation of the integral in Equation (\ref{m_recycle}) and the treatment of metallicity dependence follow the same approach as the calculation of the total supernova energy. All recycled materials are added into the cold gas component. This age-dependent mass recycling scheme is introduced due to the short timestep of the halo merger tree, and is more realistic than the commonly adopted constant recycling fraction and yield, particularly at high redshift.
\subsubsection{Reincorporation}
The ejected gas component mentioned in the previous section can be transferred back to the hot gas halo. We employ the reincorporation model proposed by \cite{2013MNRAS.431.3373H}
\begin{align}
    \Delta m_\text{reinc} &= \frac{m_\text{eject}}{t_\text{reinc}} \Delta t, \\
    t_\text{reinc} &= \gamma \frac{10^{10} \, \solarmass}{M_\text{vir}},
\end{align}
where $m_\text{eject}$ is the mass of ejected gas, and $M_\text{vir}$ is the virial mass of the friends-of-friends group. We also force the reincorporation time scale to be smaller than the halo dynamic time. The statistical analysis of \cite{2013MNRAS.431.3373H} indicates that this model provides better fit of the stellar mass functions against observations at $z \leq 3$. We set $\gamma = 18 \, \text{Gyr}$ as suggested by \cite{2013MNRAS.431.3373H}. This model is also implemented in \cite{2016MNRAS.461.1760H}, \cite{2018MNRAS.479....2C} and \cite{2018MNRAS.481.3573L}. We note that with this choice of $\gamma$, the reincorporation time scale equals the forced upper limit, i.e. the halo dynamical time, for $M_\text{vir} \lesssim 10^{12} M_\odot$ at the redshift range of interest in this study. At $z \sim 4 - 7$, the halo dynamical time is $\sim 100$ Myr. Therefore, reincorporation is more efficient in our model relative to \cite{2013MNRAS.431.3373H} at these redshifts, in particular for low mass halos. This behaviour is very different from the original model proposed by \cite{2013MNRAS.431.3373H}, and may impact on the predicted stellar mass functions at lower redshifts. We defer the exploration of this effect to later works. 

\section{Dust model and synthetic spectral energy distributions} \label{SED}
\subsection{Dust Model} \label{dust_model}
We implement the dust model proposed by \citep{2000ApJ...539..718C}. The transmission function due to the ISM is expressed by
\begin{equation}
T_\lambda(t) = 
	\begin{cases} 
		  \exp(-\tau^\text{ISM}_\lambda) & t \geq t_\text{BC} \\
		  \exp(-\tau^\text{ISM}_\lambda - \tau^\text{BC}_\lambda) & t < t_\text{BC}
	   \end{cases}.
\end{equation}
This model takes into account the relative stars-dust geometry of different stellar populations. Photons emitted by young stars are absorbed by an additional component due to the surrounding molecular cloud where the stars form. The birthcloud is assumed to have lifetime $t_\text{BC}$, and for stars whose age is older than $t_\text{BC}$, their starlight is only absorbed by the diffuse ISM dust. We fix $t_\text{BC} = 10$ Myr according to previous studies \citep{2000ApJ...539..718C,2008MNRAS.388.1595D}. The attenuation due to the birth cloud and diffuse ISM dust is described by their optical depths $\tau^\text{BC}_\lambda$ and $\tau^\text{ISM}_\lambda$ respectively, which should vary with different galaxies. In this study, we explore three different parametrisations, linked to star formation rate (SFR), dust-to-gas (DTG) ratio and gas column density (GCD). We name them as M-SFR, M-DTG and M-GCD respectively. In general, these properties are indirectly related to the dust. One dust production channel is from the ejecta of supernova \cite[e.g.][]{2018PhR...780....1D}, which is proportional to the SFR. Dust is also mixed with gas. Accordingly, they are expected to have similar properties. We will see that M-DTG and M-GCD have similar results since they primarily depend on the gas density.

\subsubsection{Star formation rate model} \label{SFR_model}
The dependence of the dust optical depths on SFR is motivated by observations of the CMRs at high redshifts, i.e the relation between UV continuum slope and UV magnitude. These observations suggest that more UV luminous galaxies have redder UV continuum slopes \citep{2012ApJ...756..164F,2014ApJ...793..115B,2014MNRAS.440.3714R}. Since brighter galaxies correspond to higher SFR, one could expect that SFR and dust content are positively correlated. Similar trends have been found in low redshift studies \citep[e.g.][]{2010MNRAS.403.1894D,2019MNRAS.485.5733Q}. Hence, we assume the following parameterisation
\begin{align}
&\Gamma_\lambda = e^{-az} \left ( \frac{\rm SFR}{100 \, \solarmassperyear}  \right)^{\gamma_\text{SFR}} \left ( \frac{\lambda}{1600\, \text{\AA}} \right )^{n}, \label{G_SFR} \\ \label{ISM_SFR}
&\tau^\text{ISM}_\lambda = \tau^\text{ISM}_{\rm SFR} \Gamma_{\lambda}, \\ \label{BC_SFR}
&\tau^\text{BC}_\lambda = \tau^\text{BC}_{\rm SFR} \Gamma_{\lambda},
\end{align}
where $\tau^\text{ISM}_{\rm SFR}$, $\tau^\text{ISM}_{\rm SFR}$, $\gamma_\text{SFR}$, $a$ and $n$ are free parameters. \cite{2019MNRAS.483.2983Y} also use a parametric model to calculate dust attenuation in their semi-analytic model. They adjusted the normalisation of the optical depth to fit the observed UV LFs at individual redshifts. Their results indicate that the normalisation depends on redshift and the trend can be fit by an exponential function. Therefore, for all the three parametrisations proposed in this work, we also include an exponential redshift dependence factor to fit the model against multiple redshifts.

\subsubsection{Dust-to-gas ratio model} \label{DTG_model}
In the literature, dust optical depths are often linked to the gas column density, which is then converted to the dust column density using the dust-to-gas (DTG) ratio \citep{2007MNRAS.375....2D,2011MNRAS.413..101G,2012MNRAS.423.1992S,2019MNRAS.483.2983Y}. In this model, optical depths are expressed by
\begin{align}
&\Gamma_\lambda = e^{-az}
\left(\frac{Z_\text{cold}}{Z_\odot} \right)^{\gamma_\text{DTG}}
\left(\frac{m_\text{cold}}{10^{10}h^{-1} \, \solarmass}\right)
\left(\frac{r_\text{disk}}{h^{-1} \, \rm kpc} \right)^{-2}
\left ( \frac{\lambda}{1600\, \text{\AA}} \right )^{n}, \label{G_DTG} \\ \label{ISM_DTG}
&\tau^\text{ISM}_\lambda = \tau^\text{ISM}_{\rm DTG} \Gamma_{\lambda}, \\ \label{BC_DTG}
&\tau^\text{BC}_\lambda = \tau^\text{BC}_{\rm DTG} \Gamma_{\lambda},
\end{align}
where $Z_\text{cold}$ is the metallicity of cold gas, $m_\text{cold}$ is the mass of cold gas, and $r_\text{disk}$ is the disk scale radius defined in Equation (\ref{r_disk}). We adopt the solar metallicity as $Z_\odot = 0.02$. Free parameters are $\tau^\text{ISM}_{\rm DTG}$, $\tau^\text{BC}_{\rm DTG}$, $\gamma_\text{DTG}$, $a$ and $n$.

\subsubsection{Gas column density model} \label{GCD_model}
We propose an additional gas mass related dust model, which is independent of the metallicity. In \citetalias{2016MNRAS.462..250M} and this work, when metals are produced by supernova explosions, we assume that they are first fully mixed with cold gas, and then ejected into the hot gas reservoir. In reality, since the materials produced by supernova have quite different initial velocities from the surrounding gas, the mixing may take some time. Thus, we provide a metallicity independent parametrisation of the dust optical depths
\begin{align}
&\Gamma_\lambda = e^{-az}
\left(\frac{m_\text{cold}}{10^{10}h^{-1} \, \solarmass}\right)^{\gamma_\text{GCD}}
\left(\frac{r_\text{disk}}{h^{-1} \, \rm kpc} \right)^{-2}
\left ( \frac{\lambda}{1600\, \text{\AA}} \right )^{n}, \label{G_GCD} \\ \label{ISM_GCD}
&\tau^\text{ISM}_\lambda = \tau^\text{ISM}_{\rm GCD} \Gamma_{\lambda}, \\ \label{BC_GCD}
&\tau^\text{BC}_\lambda = \tau^\text{BC}_{\rm GCD} \Gamma_{\lambda}.
\end{align}
There are also five free parameters in this model, i.e. $\tau^\text{ISM}_{\rm GCD}$, $\tau^\text{BC}_{\rm GCD}$, $\gamma_\text{GCD}$, $a$ and $n$. This model includes a power law scaling on the cold gas mass, unlike the M-DTG model, where the scaling is on metallicity.

\subsection{Synthetic spectral energy distributions}
The computation of galaxy spectral energy distributions (SEDs) follows standard stellar population synthesis. The luminosity of a galaxy at time $t$ can be obtained by
\begin{equation} \label{L_t}
L_\lambda(t) = \int^t_0 d\tau \int^{Z_\sub{max}}_\sub{Z_{min}} dZ \, \psi_\sub{}(t - \tau, Z)S_\lambda(\tau, Z)T_\lambda(\tau),
\end{equation}
where $\tau$ is the stellar age, $\psi(t - \tau, Z) \, d\tau dZ$ is the mass of stars formed at $t - \tau$ with an age between $\tau$ to $\tau + d\tau$ and metallicity between $Z$ to $Z + dZ$, $S_\lambda(\tau, Z)$ is the luminosity of a simple stellar population (SSP) per unit mass, and $T_\lambda(\tau)$ is the transmission function of the ISM described in the previous subsection. We generate $S_\lambda(\tau, Z)$ using \textsc{starburst99} \citep{1999ApJS..123....3L,2005ApJ...621..695V,2010ApJS..189..309L,2014ApJS..212...14L}, assuming a metallicity range from $Z = 0.001$ to $Z = 0.040$ and a \cite{2002Sci...295...82K} IMF. Nebular continuum emssions are also added using \textsc{starburst99}. To compute UV magnitudes, we apply a tophat filter centred at $\lambda = 1600 \text{\AA}$ with width $100 \text{\AA}$. UV slopes are obtained by a linear fit in the logarithmic flux space using the ten windows proposed by \cite{1994ApJ...429..582C}. However, for computational speed, we only choose five of them (including the longest wavelength window) for on-the-fly calibrations. The selected windows are given in Table \ref{windows}. The median errors from this treatment are negligible in the range of the observed CMRs.
\par
We also make a numeric approximation in order to accelerate the speed of evaluating Equation (\ref{L_t}). We first compute the intrinsic luminosity in necessary filters. The dust transmission is then applied to the luminosity of the filters using the central wavelength instead of the full SEDs. This approximation is found to have a negligible effect on the results, since all filters used in this work have a simple shape and are relatively narrow.

\begin{table}
    \centering
    \begin{tabular}{c|c}
      & Wavelength range [\AA] \\ \hline
    1 & 1342 - 1371 \\
    2 & 1562 - 1583 \\
    3 & 1866 - 1890 \\
    4 & 1930 - 1950 \\
    5 & 2400 - 2580 \\ \hline                             
    \end{tabular}
    \caption{The five windows selected from \citet{1994ApJ...429..582C} to fit UV slopes for the on-the-fly calibrations.}
    \label{windows}
\end{table}

\begin{table*}
    \caption{Summary of free galaxy and dust parameters.}
    \centering
    \resizebox{\textwidth}{!}{%
    \begin{tabular}{ccccccccccc}
    \hline \hline
    Parameter & \multicolumn{2}{c}{Section} & \multicolumn{2}{c}{Equation} & \multicolumn{6}{c}{Description} \\ \hline
    $\alpha_\text{SF}$ & \multicolumn{2}{c}{Sec.\ref{SF}} & \multicolumn{2}{c}{Eq.\ref{alpha_SF}} & \multicolumn{6}{l}{Star formation efficiency} \\
    $\Sigma_\text{SF}$ & \multicolumn{2}{c}{Sec.\ref{SF}} & \multicolumn{2}{c}{Eq.\ref{m_crit}} & \multicolumn{6}{l}{Critical mass normalisation} \\
    $\eta_0$ & \multicolumn{2}{c}{Sec.\ref{feedback}} & \multicolumn{2}{c}{Eq.\ref{eta}} & \multicolumn{6}{l}{Mass loading normalisation} \\
    $\epsilon_0$ & \multicolumn{2}{c}{Sec.\ref{feedback}} & \multicolumn{2}{c}{Eq.\ref{epsilon}} & \multicolumn{6}{l}{supernova energy coupling normalisation} \\
    $\tau^\text{ISM}_\text{SFR}$/$\tau^\text{ISM}_\text{DTG}$/$\tau^\text{ISM}_\text{GCD}$ & \multicolumn{2}{c}{Sec.\ref{SFR_model}/Sec.\ref{DTG_model}/Sec.\ref{GCD_model}} & \multicolumn{2}{c}{Eq.\ref{ISM_SFR}/Eq.\ref{ISM_DTG}/Eq.\ref{ISM_GCD}} & \multicolumn{6}{l}{Dust optical depth normalisation of ISM} \\
    $\tau^\text{BC}_\text{SFR}$/$\tau^\text{BC}_\text{DTG}$/$\tau^\text{BC}_\text{GCD}$ & \multicolumn{2}{c}{Sec.\ref{SFR_model}/Sec.\ref{DTG_model}/Sec.\ref{GCD_model}} & \multicolumn{2}{c}{Eq.\ref{BC_SFR}/Eq.\ref{BC_DTG}/Eq.\ref{BC_GCD}} & \multicolumn{6}{l}{Dust optical depth normalisation of BC} \\
    $\gamma_\text{SFR}$/$\gamma_\text{DTG}$/$\gamma_\text{GCD}$ & \multicolumn{2}{c}{Sec.\ref{SFR_model}/Sec.\ref{DTG_model}/Sec.\ref{GCD_model}} & \multicolumn{2}{c}{Eq.\ref{G_SFR}/Eq.\ref{G_DTG}/Eq.\ref{G_GCD}} & \multicolumn{6}{l}{Dust optical depth slope of galaxy property} \\
    $n$ & \multicolumn{2}{c}{Sec.\ref{SFR_model}/Sec.\ref{DTG_model}/Sec.\ref{GCD_model}} & \multicolumn{2}{c}{Eq.\ref{G_SFR}/Eq.\ref{G_DTG}/Eq.\ref{G_GCD}} & \multicolumn{6}{l}{Reddening slope} \\
    $a$ & \multicolumn{2}{c}{Sec.\ref{SFR_model}/Sec.\ref{DTG_model}/Sec.\ref{GCD_model}} & \multicolumn{2}{c}{Eq.\ref{G_SFR}/Eq.\ref{G_DTG}/Eq.\ref{G_GCD}} & \multicolumn{6}{l}{Dust optical depth redshift dependence} \\ \hline
    Parameter & Prior scale & \multicolumn{3}{c}{Prior range} & \multicolumn{3}{c}{Best-fit  $^\text{a}$} & \multicolumn{3}{c}{16/84-th percentiles  $^\text{b}$} \\
     &  & M-SFR & M-DTG & M-GCD & M-SFR & M-DTG & M-GCD & M-SFR & M-DTG & M-GCD \\ \hline
    $\alpha_\text{SF}$ & log & [0.005, 0.2] & [0.05, 0.18] & [0.04, 0.08] & 0.10 & 0.10 & 0.05 & [0.08, 0.13] & [0.10, 0.11] & [0.05, 0.07] \\
    $\Sigma_\text{SF}$ & log & [0.1, 0.8] & [0.001, 0.25] & [0.05, 0.25] & 0.19 & 0.01 & 0.16 & [0.21, 0.42] & [0.007, 0.06] & [0.14, 0.19] \\
    $\eta_0$ & log & [2.0, 12.0] & [2.0, 15.0] & [3.5, 7.5] & 4.6 & 7.0 & 6.4 & [4.0, 7.8] & [6.6, 7.9] & [4.9, 6.1] \\
    $\epsilon_0$ & log & [0.35, 0.65] & [0.8, 2.2] & [1.0, 1.7] & 0.5 & 1.5 & 1.3 & [0.4, 0.6] & [1.5, 1.7] & [1.3, 1.5] \\
    $\tau^\text{ISM}_\text{SFR}$/$\tau^\text{ISM}_\text{DTG}$/$\tau^\text{ISM}_\text{GCD}$ & linear & [0.5, 2.4] & [0.0, 50.0] & [2.0, 8.0] & 1.7 & 13.5 & 3.7 & [1.4, 1.7] & [9.9, 17.0] & [3.5, 5.3] \\
    $\tau^\text{BC}_\text{SFR}$/$\tau^\text{BC}_\text{DTG}$/$\tau^\text{BC}_\text{GCD}$ & linear & [2.0, 10.0] & [0.0, 1000.0] & [25.0, 140.0] & 2.5 & 381.3 & 69.7 & [3.9, 6.6] & [225.1, 476.1] & [60.4, 91.0] \\
    $\gamma_\text{SFR}$/$\gamma_\text{DTG}$/$\gamma_\text{GCD}$ & linear & [0.0, 0.6] & [0.4, 2.2] & [1.3, 1.7] & 0.19 & 1.20 & 1.48 & [0.23, 0.32] & [1.05, 1.38] & [1.44, 1.52] \\
    $n$ & linear & [-1.00, -0.25] & [-2.5, -0.8] & [-1.6, -1.0] & -0.3 & -1.6 & -1.3 & [-0.5, -0.3] & [-1.7, -1.5] & [-1.4, -1.2] \\
    $a$ & linear & [0.00, 0.15] & [0.10, 0.65] & [0.20, 0.55] & 0.04 & 0.34 & 0.39 & [0.02, 0.07] & [0.25, 0.37] & [0.36, 0.42] \\ \hline
    \end{tabular}
    }%
    \begin{tablenotes}
        \item $^\text{a}$Sample point that has the highest posterior distribution value are chosen to be the best-fit values.
        \item $^\text{b}$These are the 16-th and 84-th percentiles of the marginalised distributions.
    \end{tablenotes}
    \label{params}
\end{table*}

\begin{figure*}
 \includegraphics[width=\textwidth]{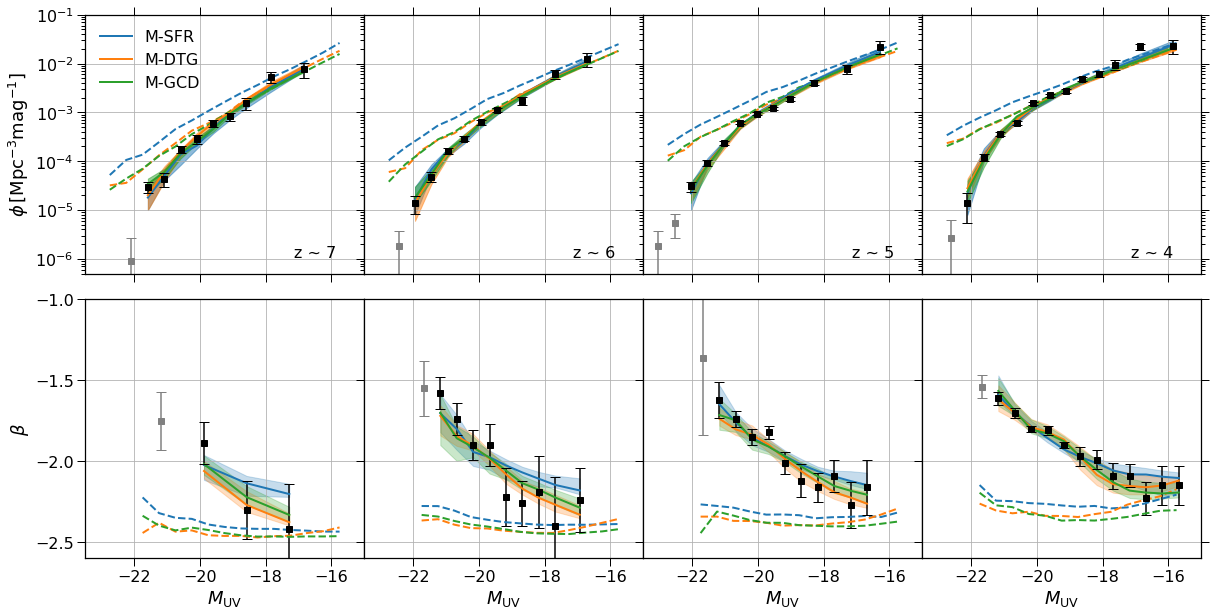}
 \caption{Best-fit luminosity functions (LFs) and colour-magnitude relations (CMRs). Solid blue, orange and green lines are the results of M-SFR (Section \ref{SFR_model}), M-DTG (Section \ref{DTG_model}), M-GCD (Section \ref{GCD_model}) respectively. Shaded regions illustrate the $1\sigma$ (68 \%) range of the posterior distributions. Dashed lines are the corresponding dust-unattenuated properties. Black points with errorbars are the observational data used in the calibration, which are from \citet{2015ApJ...803...34B} and \citet{2014ApJ...793..115B} for the LFs and CMRs respectively. Grey data points are also from these observations but are not used in the calibration due to the limit of the simulation box size.}
 \label{best-fit}
\end{figure*}

\begin{figure*}
 \includegraphics[width=\textwidth]{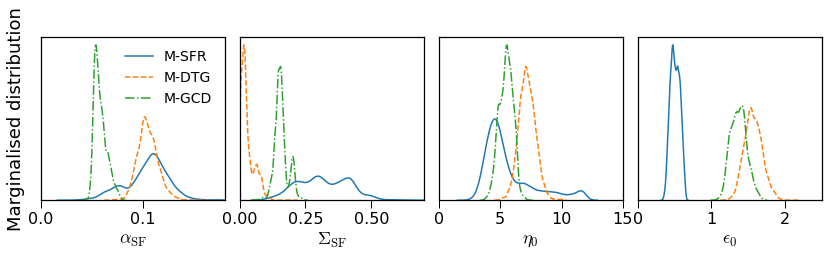}
 \caption{Comparison of the marginalised distributions of galaxy formation parameters among the three different dust models. These parameters are the star formation efficiency $\alpha_\text{SF}$ (Equation \ref{alpha_SF}), the normalisation of the critical mass  $\Sigma_\text{SF}$ (Equation \ref{m_crit}), the mass loading factor $\eta_0$ (Equation \ref{eta}) and the supernova energy coupling efficiency $\epsilon_0$ (Equation \ref{epsilon}). The three dust models labelled as M-SFR, M-DTG and M-GCD are described in Section \ref{SFR_model}, \ref{DTG_model} and \ref{GCD_model}, and the corresponding optical depths in the three models are linked to the star formation rate (SFR), dust to gas ratio (DTG) and gas column density (GCD) respectively. The y-axes show the probability distributions in a linear scale.}
 \label{pdf_cp}
\end{figure*}

\section{Calibration} \label{calib}
An essential part of this work is to determine the free parameters in both the galaxy formation and dust attenuation models introduced in the previous sections. We carry out a Bayesian analysis on these parameters, and use observed UV LFs and CMRs at $z \sim 4 - 7$ as constraints.
\par
A key goal of a Bayesian analysis is to estimate the posterior distribution of model parameters, which is non-trivial for high dimensional spaces. \cite{2008MNRAS.384.1414K} and \cite{2009MNRAS.396..535H} first applied the Markov chain Monte Carlo (MCMC) method to sample the parameter space of semi-analytic models. This approach has been implemented by several subsequent studies \citep{2013MNRAS.431.3373H,2013MNRAS.428.2001M,2015MNRAS.451.2663H}. However, the MCMC method has several drawbacks. Firstly, it requires additional evaluations of the model to ensure the final sample reaches a stationary distribution, and it is generally difficult to determine whether a Monte Carlo chain has fully converged \cite[see][for a review]{doi:10.1080/01621459.1996.10476956}. Moreover, without special treatments, MCMC samplers can encounter difficulties in approaching a stationary distribution when the parameter space contains isolated modes (which is the case in this study), since random walkers can be trapped by a local minimum and fail to jump to other modes \cite[e.g][]{Neal1996}. A possible improvement to handle multimodal distributions for MCMC methods can be found in \cite{2005PCCP....7.3910E}.
\par
In this work, in order to achieve higher sampling efficiency and obtain more stable results on multimodal parameter spaces, we utilise the multimodal nested sampling introduced by \cite{2009MNRAS.398.1601F} to estimate the posterior distributions. This algorithm is found to be a competitive alternative to MCMC methods, and addresses the issues mentioned above to some extent. The nested sampling was designed to evaluate the Bayesian evidence \citep{2004AIPC..735..395S}. However, the output samples produced by the algorithm can also be used to estimate posterior distributions, which is equivalent to the MCMC method. In difference from MCMC methods, no burn-in phase is required in this algorithm. The stopping criterion of the nested sampling is based on an estimated error of the resulting value of the Bayesian evidence, which is also proposed by \cite{2004AIPC..735..395S}. The sampling efficiency of the original algorithm is improved by \cite{2009MNRAS.398.1601F}, who use the information of existing sample points to approximate the iso-likelihood surfaces in the parameter space as hyper ellipsoids \citep[see also][]{2006ApJ...638L..51M}. Secondly, the algorithm includes a special treatment for multimodal problems. Again using the information of existing sample points, it applies a clustering algorithm to detect multiple modes and splits the parameter space \cite[see also][]{2007MNRAS.378.1365S}. This approach has been tested against toy models which contain several equally-high peaks, and is found to have good performance. The reader is referred to \cite{2008MNRAS.384..449F}, \cite{2009MNRAS.398.1601F} and references therein for a detailed description of the algorithm. A comparison between the nested sampling and the MCMC method can found be in \cite{2019arXiv190402180S}.
\par
The Bayesian posterior distribution is comprised of the likelihood and prior distributions of each free model parameter. We construct the log-likelihood as
\begin{equation}
\begin{split}
\ln \mathcal{L} = &-\frac{1}{2} \sum\limits_i \left [\frac{(n^\sub{obs}_i - n^\sub{model}_i)^2}{\sigma^2_\sub{LF, i} } + \ln (2 \pi \sigma^2_\sub{LF, i}) \right] \\
                  &-\frac{1}{2} \sum\limits_i \left [\frac{(\beta^\sub{obs}_i - \beta^\sub{model}_i)^2}{\sigma^2_\sub{CMR, i} } + \ln (2 \pi \sigma^2_\sub{CMR, i}) \right].
\end{split}
\end{equation}
Observational data of LFs ($n^\sub{obs}_i$, $\sigma^2_\sub{LF, i}$) and CMRs ($\beta^\sub{obs}_i$, $\sigma^2_\sub{CMR, i}$) are taken from \cite{2015ApJ...803...34B} and the biweight mean measurements of \cite{2014ApJ...793..115B} respectively. The LFs are defined by the co-moving number density. We convert the dimensionless Hubble constant from $h = 0.7$ to $h = 0.678$ for these observations in order to be consistent with our model. Due to the limited size of the simulation box, the model is unable to probe the full range of the LFs and CMRs. Therefore, for each LF and CMR bin, we use the observed LF to estimate an expected number of galaxies in the simulation box, and drop the bin if the number is less than five and twenty for the LF and CMR respectively.
\par
The model parameters are from both the semi-analytic model and the dust relations introduced in Section \ref{dust_model}. We focus on four galaxy formation parameters: the star formation efficiency $\alpha_\text{SF}$, normalisation of the critical mass $\Sigma_\text{SF}$, mass loading factor $\eta_0$, and supernova energy coupling efficiency $\epsilon_0$. Their prior distributions are chosen to be uniform in logarithmic space. Three different dust models were introduced in Section \ref{dust_model}. Each of them has five free parameters. We adopt uniform priors in linear space for them.
\par
The prior ranges of these model parameters are used in the initialisation of nested sampling, and they are chosen based on several experiments. We first run the sampler in a very large parameter space and find the high probability regions. We then shrink the prior ranges accordingly, keeping the posterior distribution at the bounds negligible compared with the high probability regions. There is an exception for the mass loading factor $\eta_0$ in the M-SFR model, which is found to have no upper limit. However, this will not affect our main results, since the energy coupling efficiency $\epsilon_0$ puts physical upper limit on the strength of the supernova feedback, and this parameter is constrained. This approach of choosing the prior ranges allows the sampler to spend more time on the high probability regions and improves the sampling efficiency. A summary of all model parameters and their prior ranges can be found in Table \ref{params}.
\par
In practice, we utilise a modified verion of the open source Python package \textsc{nestle}\footnote[1]{\label{nestle}https://github.com/kbarbary/nestle. See https://github.com/smutch/nestle for the modified version.}, which implements the algorithm, and couple it with the \textsc{meraxes} Python interface \textsc{mhysa} (Mutch, in prep.). We set the number of active points to be 300 for the sampler. The stop criterion follows the remaining Bayesian evidence approach suggested by \cite{2004AIPC..735..395S}. The algorithm terminates when the logarithmic change due to the remaining Bayesian evidence is below one, and the convergence requires evaluating the model for 50,000 - 100,000 times.

\begin{figure*}
 \includegraphics[width=0.75\textwidth]{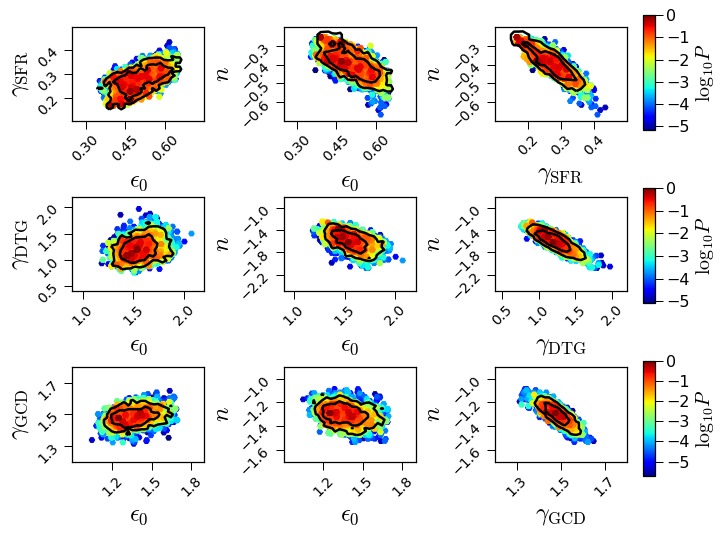}
 \caption{Correlations among the supernova energy coupling efficiency $\epsilon_0$, galaxy property scaling of the dust relation $\gamma_\text{SFR,\,DTG,\,GCD}$ and the reddening slope $n$. In each panel, solid back lines are the 68\% and 95\% contours of the two parameter marginalised distributions. Colour points indicate the values of the corresponding posterior distributions, and their maximum is normalised to unity. From top to bottom, rows correspond to the dust attenuation model of M-SFR (Section \ref{SFR_model}), M-DTG (Section \ref{DTG_model}), M-GCD (Section \ref{GCD_model}) respectively. The posterior distributions of all parameters for the three models can be seen in Figures \ref{pdf_SFR}, \ref{pdf_DTG} and \ref{pdf_GCD}.}
 \label{pdf_short}
\end{figure*}

\begin{figure}
 \includegraphics[width=\columnwidth]{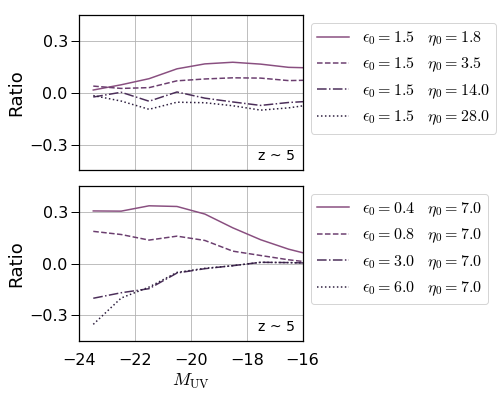}
 \caption{Effects of varying the mass loading factor $\eta_0$ and the supernova energy coupling efficiency $\epsilon_0$ on the intrinsic UV luminosity function. The y-axes show the ratio of the logarithmic luminosity functions between the model variants and the best-fit M-DTG models. In the upper panel, we vary the mass loading factor $\eta_0$ at fixed $\epsilon_0$, while in the lower panel, we fix $\epsilon_0$ and change $\eta_0$.}
 \label{sn_effect}
\end{figure}

\begin{figure}
 \includegraphics[width=\columnwidth]{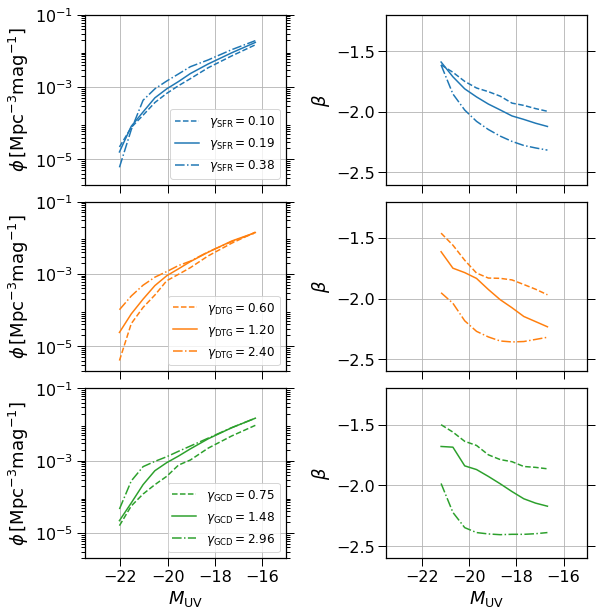}
 \caption{Effects of varying the galaxy property scaling of the dust relation $\gamma_\text{SFR,\,DTG,\,GCD}$ on the dust-attenuated UV luminosity functions and colour magnitude relations. In each panel, solid lines correspond to the results of the best-fit models as shown in Figure \ref{best-fit}. Dashed and dot dashed lines show the results of the model variants, in which $\gamma_\text{SFR,\,DTG,\,GCD}$ is changed by a factor of two. From top to bottom, rows correspond to the dust attenuation model of M-SFR (Section \ref{SFR_model}), M-DTG (Section \ref{DTG_model}), M-GCD (Section \ref{GCD_model}) respectively.}
 \label{gamma}
\end{figure}

\section{fitting results} \label{res}
For the three different dust models, we obtain $5,000$ - $6,000$ sample points from the nested sampling algorithm, which describe the posterior distributions of both galaxy and dust parameters. The point that has highest value of the posterior distribution is chosen to be the best-fit result. The best-fit parameter values are listed in Table \ref{params}, and the corresponding LFs and CMRs are shown in Figure \ref{best-fit} for each dust model. The three models all fit the observational data extremely well. In Figures \ref{pdf_SFR}, \ref{pdf_DTG} and \ref{pdf_GCD}, we show the posterior distributions of M-SFR, M-DTG and M-GCD respectively. In plotting these figures, we adopt a similar approach with \cite{2009MNRAS.396..535H} and \cite{2013MNRAS.431.3373H}, i.e. using contours to show the marginalised distributions and colours to indicate the values of the whole posterior distributions. For the M-SFR model, it can be seen from Figure \ref{pdf_cp} that the marginalised distribution of the mass loading factor $\eta_0$ extends to large values, which means that this parameter is less constrained. On the other hand, all parameters are well constrained for the other two models.
\par
An interesting finding is that the derived galaxy formation parameters preferred by these three dust models are quite different. Figure \ref{pdf_cp} illustrates a comparison of the marginalised distributions for the four galaxy formation parameters. We found that M-DTG and M-GCD suggest similar mass loading factor and supernova energy coupling efficiency. However, M-DTG shows evidence of a more active star formation scenario, with higher star formation efficiency and lower normalisation of the critical mass. For M-SFR, the marginalised distribution of the star formation efficiency overlaps with that of M-DTG. However, M-SFR requires much smaller supernova energy coupling efficiency. Moreover, differences can also be found in the two parameter correlations between the galaxy formation parameters for the three different models. For instance, in the third row and first column of Figure \ref{pdf_GCD}, M-GCD shows a strong correlation between the star formation efficiency $\alpha_\text{SF}$ and the mass loading factor $\eta_0$. However, this correlation cannot be found the in the other two models. The variation among the posterior distributions of the three models implies that these free parameters fit the observational data in a very complex way and the constraints on them depend on the assumptions used to model the dust attenuation.
\par
By comparing the posterior distributions of the three dust models, we find similar correlations among the parameters of the supernova feedback, galaxy property scaling of the dust relation and reddening slope. We demonstrate this in Figure \ref{pdf_short}. It can be seen that the supernova energy coupling efficiency is positively and inversely correlated with $\gamma_\text{SFR,\,DTG,\,GCD}$ and the reddening slope $n$, respectively. While the trends are the weakest for the M-GCD model, the correlation between $\gamma_\text{SFR,\,DTG,\,GCD}$ and $n$ is obvious for all the three models. Similar correlations are also found for the mass loading factor $\eta_0$. The reader is referred to Figures \ref{pdf_SFR}, \ref{pdf_DTG} and \ref{pdf_GCD} for the two parameter marginalised distributions of all model parameters. The dependence between the galaxy formation and dust parameters is important, since it suggests that the observations can put constraints on intrinsic or dust-unattenauted galaxy properties despite the degrees of freedom in the dust models.
\par
In order to understand the correlations mentioned above, we plot the intrinsic LFs and CMRs for the three best-fit models in Figure \ref{best-fit}. They are shown as dashed lines. It can be seen that the LFs of the best-fit M-SFR is roughly a factor of two higher than for the other two models. We note that the intrinsic luminosity functions are more sensitive to feedback processes rather than the star formation law due to self-regulation \cite[e.g.][]{2010MNRAS.402.1536S,2011MNRAS.416.1566L}. The supernova coupling efficiency of the best-fit M-SFR model is much smaller than the other two models, which is likely to be the main reason for the difference in the intrinsic LFs, irrespective of those star formation parameters. We examine the effect of supernova feedback in Figure \ref{sn_effect}. For the upper panel, we vary the mass loading factor $\eta_0$ at fixed supernova energy coupling efficiency $\epsilon_0$ for the best-fit M-DTG model, and compare the resulting LFs with the best-fit results. The y-axis shows the ratio of the logarithmic intrinsic LFs. We find that the number density at fixed UV magnitude decreases with increasing $\eta_0$. Since the energy coupling efficiency puts an upper limit on the reheated mass (see Equation \ref{m_reheat}), the change in the LFs is smaller at larger $\eta_0$. The results of varying $\epsilon_0$ at fixed $\eta_0$ are shown in the lower panel of Figure \ref{sn_effect}. While higher $\epsilon_0$ decreases the LFs, the effect is found to be more significant at the bright end. Since the energy coupling efficiency is assumed to scale as a power law of the maximum circular velocity (see Equation \ref{epsilon}), the efficiency can easily reach the maximum value of unity for small galaxies. The median $M_\text{UV}$ - $V_\text{max}$ relation of the best-fit M-DTG model indicates that the energy coupling efficiency becomes unity at an intrinsic magnitude $M_\text{UV} \sim -18$ with $\epsilon_0 = 1.5$. Therefore, the major effect of increasing $\epsilon_0$ is to allow more gas to be reheated in galaxies hosted by more massive halos. This explains why this parameter has larger impact at the bright end of the intrinsic LFs. Overall, the above discussion implies that supernova feedback plays an important role in regulating the intrinsic LFs.
\par
We next investigate the effect of the galaxy property scaling of the dust relation. Figure \ref{gamma} shows the resulting dust-attenuated LFs and CMRs when $\gamma_\text{SFR,DTG,GCD}$ is changed by a factor of two. It can be seen that the shape of both the LFs and CMRs are quite sensitive to this parameter. Furthermore, since the dust optical depths are assumed to depend on different galaxy properties, this parameter changes the shape of the LFs and CMRs in different ways.
\par
Combining the discussions of the supernova feedback parameters and $\gamma_\text{SFR,DTG,GCD}$ above, we provide an explanation of the correlations between the model parameters in Figure \ref{pdf_short}. In our dust models, the effective UV optical depth is a function of the galaxy property scaling $\gamma_\text{SFR,DTG,GCD}$ and the optical depth normalisations of both the ISM and BC. The galaxy property scaling has a direct impact on the shape of the dust-attenuated LFs and CMRs, and this single parameter is required to satisfy two shapes. Hence, the effective UV optical depth is very sensitive to $\gamma_\text{SFR,DTG,GCD}$. The effective optical depth should be degenerate with the intrinsic UV LFs, which are primarily controlled by the supernova feedback parameters. These imply that both $\eta_0$ and $\epsilon_0$ should be correlated with $\gamma_\text{SFR,DTG,GCD}$. On the other hand, the observed UV continuum slope $\beta$ depends on the reddening curve, which is assumed to be a power law of wavelength. Since there is a natural degeneracy between the power law slope and the normalisation, the reddening slope $n$ should be degenerate with the effective UV optical depth, and therefore $\gamma_\text{SFR,DTG,GCD}$. The dependence between the supernova feedback parameters and the reddening slope can be derived from the above two correlations. Figure \ref{gamma} shows that the galaxy property scaling on the SFR changes the dust-attenuated LFs and CMRs differently from the other two models, which may explain why the best-fit M-SFR model requires much weaker supernova feedback. This also explains the shallower reddening slope required by the best-fit M-SFR model.
\par
\begin{figure}
 \includegraphics[width=0.9\columnwidth]{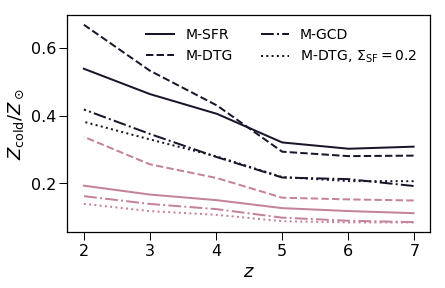}
 \caption{Redshift evolution of the mass metallicity relation. The y-axis represents the cold gas metallicity, with $Z_\odot = 0.02$. Dark and light lines correspond to the metallicity at different stellar mass bins, $10^{8} \solarmass < M_* < 10^{8.5} \solarmass$ and $10^{9} \solarmass < M_* < 10^{9.5} \solarmass$ respectively. Solid, dashed and dot dashed lines show the best-fit results of M-SFR (Section \ref{SFR_model}), M-DTG (Section \ref{DTG_model}) and M-GCD (Section \ref{GCD_model}) respectively. The metallicity only depends on the galaxy formation parameters of these models, which are listed in Table \ref{params}. The dotted lines correspond to the results of a model variant for which the normalisation of critical mass is set to be $\Sigma_\text{SF}= 0.2$, while other parameters remain the same with the best-fit M-DTG. This variant model illustrates that $\Sigma_\text{SF}$ is a main driver of the cold gas metallicity.}
 \label{metals}
\end{figure}

In addition, we contrast the best-fit models for M-DTG and M-GCD. Whilst their intrinsic LFs and CMRs are almost the same, a difference is found in the metallicity. Figure \ref{metals} illustrates the cold gas metallicity at two stellar mass bins as a function of redshift for all best-fit models. It is clear that the cold gas is more metal enriched in the best-fit M-DTG than in M-GCD. We identify that the normalisation of the critical mass $\Sigma_\text{SF}$ is the primary driver for the difference since both best-fit models have similar parameters of supernova feedback. To confirm this, we run a model variant, setting $\Sigma_\text{SF} = 0.2$, with other parameters being the same with the best-fit M-DTG. The resulting metallicity is also shown in Figure \ref{metals}, which is similar to that of the best-fit M-GCD. This finding is unsurprising since the star formation law affects gas fraction and therefore metallicity \cite[e.g.][]{2010MNRAS.402.1536S,2011MNRAS.416.1566L}.  In addition, from Figure \ref{metals}, it is worth noting that the metallicity evolves with redshift in our model, with higher metallicity at lower redshifts. This is expected due to the explicit redshift dependence on the mass loading factor, which is motivated by previous studies \citep{2015MNRAS.454.2691M,2016MNRAS.461.1760H,2018MNRAS.481..954C}.

\begin{figure*}
 \includegraphics[width=0.9\textwidth]{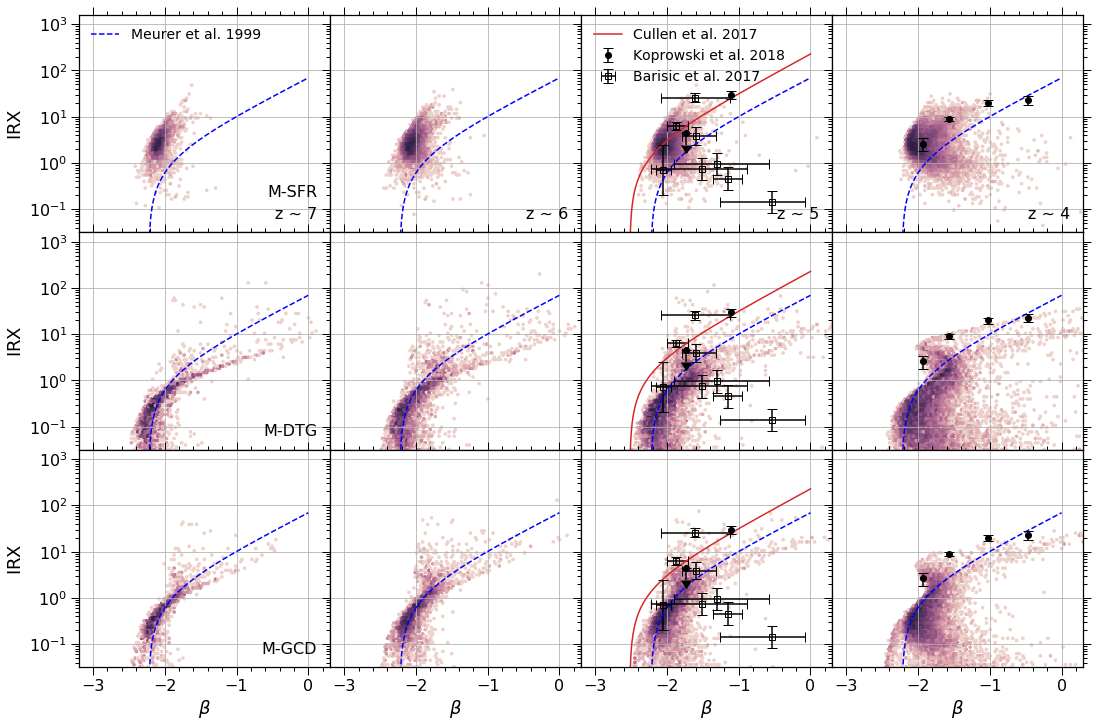}
 \caption{Predicted infrared excess (IRX) - UV continuum slope $\beta$ relations. From top to bottom, rows show the results of the best-fit M-SFR (Section \ref{SFR_model}), M-DTG (Section \ref{DTG_model}) and M-GCD (Section \ref{GCD_model}) models. We only show model galaxies with stellar mass greater than $10^8 \solarmass$. The dust optical depths in the three models are linked to the star formation rate (SFR), dust to gas ratio (DTG) and gas column density (GCD) respectively. The relations are represented by purple density plots. The best-fit parameters of these models can be seen from Table \ref{params}. IRX is computed by energy balance arguments. Columns show the results at different redshifts. Blue dashed lines are the widely used \citet{1999ApJ...521...64M} relation. Red lines show the results from \citet{2017MNRAS.470.3006C}, which are based on the post-process of the FiBY hydrodynamic simulation \citep{2013MNRAS.428.1857J,2015MNRAS.451.2544P}. Black circles with error bars are stacking measurements of Lyman-break galaxies (LBGs) from \citet{2018MNRAS.479.4355K}. Individual source measurements from \citet{2017ApJ...845...41B} are shown as empty squares.}
 \label{plot_IRX}
\end{figure*}

\begin{figure*}
 \includegraphics[width=0.8\textwidth]{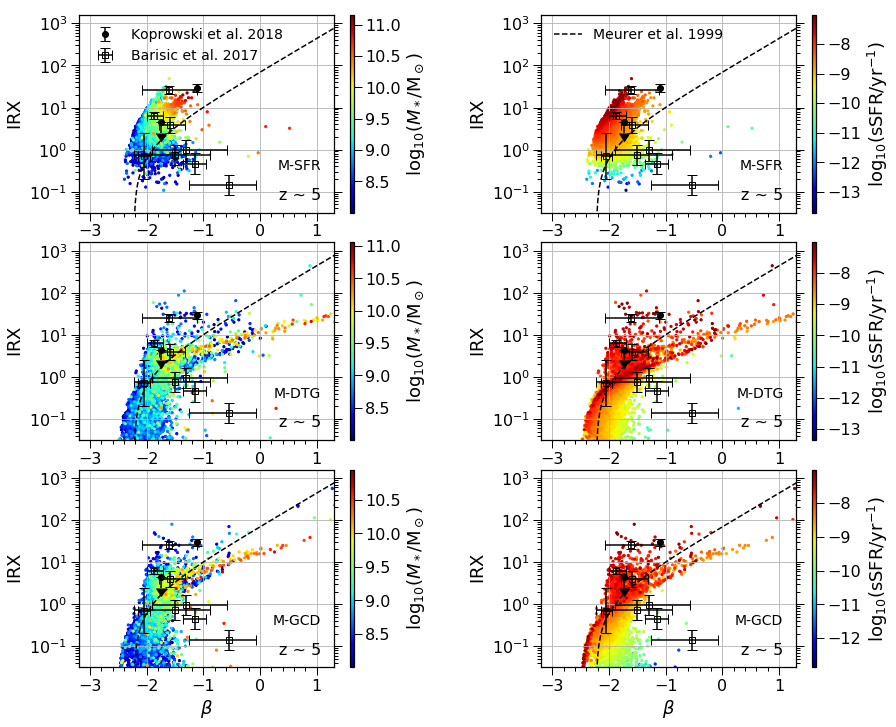}
 \caption{Predicted infrared excess (IRX) - UV continuum slope $\beta$ relations as functions of stellar mass (left panels) and specific star formation rate (sSFR) (right panels) at $z \sim 5$. We only show model galaxies with stellar mass greater than $10^8 \solarmass$. From top to bottom, rows show the results of the best-fit M-SFR (Section \ref{SFR_model}), M-DTG (Section \ref{DTG_model}), M-GCD (Section \ref{GCD_model}). The dust optical depths in the three models are linked to the star formation rate (SFR), dust to gas ratio (DTG) and gas column density (GCD) respectively. Black dashed lines show the relation measured by \citet{1999ApJ...521...64M}. Black circles and empty squares with error bars are stacking and individual measurements from \citet{2018MNRAS.479.4355K} and \citet{2017ApJ...845...41B} respectively.}
 \label{IRX_scatter}
\end{figure*}

\section{Infrared excess to UV continuum slope relations} \label{IRX}
As demonstrated in previous sections, by simultaneously fitting our galaxy formation and dust models to the observed UV LFs and CMRs, we are able to obtain constraints on both the dust attenuation in the UV band and the reddening. This allows estimates of the infrared luminosity $F_\text{IR}$ and therefore the infrared excess (IRX) using energy balance arguments, i.e.
\begin{align}
    & F_\text{IR} = \int^\infty_{912 \text{\AA}} \left( L_\lambda - L^\text{intrinsic}_\lambda \right) \, d\lambda \\
    & F_\text{UV} = \lambda L_\lambda \big\rvert_{\lambda = 1600 \text{\AA}} \\
    & \text{IRX} = \frac{F_\text{IR}}{F_\text{UV}}
\end{align}
We compute the IRX for galaxies in the best-fit models of M-SFR, M-DTG and M-GCD. The resulting IRX - $\beta$ relations for galaxies with stellar mass greater than $10^8 \solarmass$ are shown in Figure \ref{plot_IRX} with several observations for comparison. Taking into account intrinsic scatter in the relations, our results cover the observations from \cite{2018MNRAS.479.4355K}, who performed a stacking analysis of 4209 Lyman-break galaxies (LBGs) at $3 \lesssim z \lesssim 5$, and individual detection from \cite{2017ApJ...845...41B}. We also compare our predictions with the relations calibrated by \cite{1999ApJ...521...64M} using local starburst galaxies. The \cite{1999ApJ...521...64M} relation is frequently used to correct dust extinction in both observational and theoretical studies at high redshifts \citep[e.g.][]{2014MNRAS.444.2960D,2015ApJ...803...34B,2015ApJ...813...21M,2016MNRAS.462..235L,2018PASJ...70S..11H}. It can be seen from Figure \ref{plot_IRX} that the best-fit M-SFR predicts higher IRX than the \cite{1999ApJ...521...64M} relation at fixed $\beta$, while the other two best-fit models suggest lower IRX. Thus, our models indicate dust extinction that differs from the \cite{1999ApJ...521...64M} relation, which implies that a direct application of the relation at high redshifts may lead to systematic errors on the dust corrections. We will discuss the resulting uncertainties on estimations of the cosmic SFRD in Section \ref{CSFD}.
\subsection{Reddening slope}
The best-fit models for M-SFR, M-DTG and M-GCD have quite different reddening slopes $n$, which can be read from Table \ref{params}. The best-fit M-SFR model has the shallowest slope of $n = -0.3$, while much steeper slopes are found for the best-fit M-DTG and M-GCD, with $n =-1.6$ and $-1.3$ respectively. This difference is directly reflected on the IRX - $\beta$ plane. In Figure \ref{IRX}, the best-fit M-DTG and M-GCD show a shallower IRX - $\beta$ relation at redder UV slope regime. Similar disagreements can also be found from other studies. For example, \cite{2017MNRAS.470.3006C} post-processed the outputs of the FiBY hydrodynamic simulation \citep{2013MNRAS.428.1857J,2015MNRAS.451.2544P}. They propose a similar dust model to the present work, linking the dust optical depths to the logarithmic stellar mass. The free parameters in their model are adjusted to fit the observed LFs and CMRs from \cite{2014MNRAS.440.3714R} at $z \sim 5$. They suggest $n = -0.55^{+0.25}_{-0.15}$. We plot their results as solid red lines in Figure \ref{plot_IRX}, which is more consistent with the best-fit M-SFR than our other models. On the other hand, \cite{2016MNRAS.462.3130M} also post-processed a hydrodynamic simulation by \cite{2010MNRAS.407.1003M}, and coupled it with an dust evolution model. Their results reproduce the observed LFs of \cite{2015ApJ...803...34B} and CMRs of \cite{2014ApJ...793..115B} at $z \sim 5 - 8$ when using a SMC-like extinction curve. The slope of the SMC-like extinction curve is steeper, and is similar to those of the best-fit M-DTG and M-GCD. Since all our models can well reproduce observed LFs and CMRs, we cannot draw any firm conclusions on the reddening slope. Instead, we treat this as systematic uncertainties arising due to different assumptions in the dust models.
\subsection{Intrinsic scatter}
At $z \gtrsim 3$, observations show considerable scatter in the IRX - $\beta$ plane \citep[e.g.][]{2015Natur.522..455C,2016A&A...587A.122A,2016ApJ...833...72B,2017ApJ...845...41B,2017MNRAS.472..483F,2018MNRAS.479.4355K}, which might be explained by the large intrinsic scatter in our predicted relations. Hence, it is instructive to examine the main drivers of the scatter. We first notice that from Figure \ref{plot_IRX}, low IRX galaxies vanish in the best-fit M-SFR model. This is due to the nature of our star formation prescription (see Section \ref{SF}). The SFR of galaxies whose cold gas mass is below the critical mass is zero. Accordingly, in the M-SFR model, the dust optical depths of these galaxies are also zero, which results in the disappearance of the IRX. This unrealistic feature shows the limitations of this model.
\par
In left and right panels of Figure \ref{IRX_scatter}, we illustrate the IRX - $\beta$ relations at $z \sim 5$ as functions of stellar mass and specific star formation rate (sSFR) respectively. The relations at other redshifts show similar trends. For the stellar mass case, we see that massive galaxies form a tight correlation between IRX and $\beta$ in the high IRX and red $\beta$ regions. The trend that more massive galaxies have higher IRX is also observed by \cite{2016A&A...587A.122A} and \cite{2017MNRAS.472..483F}. However, we also find several larger stellar mass galaxies which have lower IRX and redder $\beta$. They might explain some of the outliers individually detected by \cite{2017ApJ...845...41B}, as shown in Figure \ref{IRX_scatter}. On the other hand, the right panels show that the scatter of the IRX - $\beta$ relation is tightly correlated with sSFR. At fixed IRX, redder galaxies typically have lower sSFR. This trend is consistent with other theoretical studies \citep{2017MNRAS.472.2315P,2017ApJ...840...15S,2018MNRAS.474.1718N,2019arXiv190101747C}. In addition, it is worth noting that although the dust optical depths are related to different galaxy properties for the three dust models, we find similar dependence of the scatter in the IRX - $\beta$ plane on both stellar mass and sSFR.

\begin{figure}
 \includegraphics[width=\columnwidth]{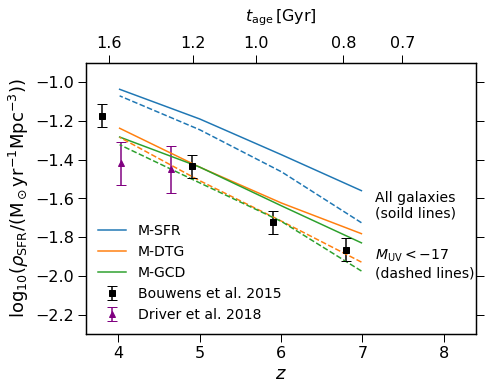}
 \caption{Predicted cosmic star formation rate density (SFRD) at $z \sim 4 - 7$. Blue, orange and green lines are estimated from the best-fit M-SFR (Section \ref{SFR_model}), M-DTG (Section \ref{DTG_model}), M-GCD (Section \ref{GCD_model}) respectively. The dust optical depths in the three models are linked to the star formation rate (SFR), dust to gas ratio (DTG) and gas column density (GCD). Solid lines are the SFRD of all model galaxies, while the results with a magnitude cut $M_\text{UV} < -17$ are shown as dashed lines. Black data points are observations from \citet{2015ApJ...803...34B}. Their dust corrections are derived by the colour magnitude relations (CMRs) of \citet{2014ApJ...793..115B} and the \citet{1999ApJ...521...64M} relation. Purple trangles with error bars show the measurements of \citet{2018MNRAS.475.2891D} using the energy balance SED-fitting code \textsc{magphys} \citep{2008MNRAS.388.1595D}.}  
 \label{plot_CSFD}
\end{figure}

\section{Cosmic star formation rate density} \label{CSFD}
Dust corrections are typically required for the conversion between the UV luminosity and the SFR. As mentioned, in high redshift observational studies of SFR, the \cite{1999ApJ...521...64M} relation is widely used, though it is calibrated against local galaxies. The previous section has shown that the dust extinction predicted by our models, which reproduce both LFs and CMRs at $z \sim 4 - 7$, is rather different from the \cite{1999ApJ...521...64M} relation. In principle, we could derive similar relations based on our results to be used by other studies to perform the dust corrections. However, by using such relations, we should be able to recover the SFR functions of our models given the LFs. Therefore, we directly present the predicted SFRs. Furthermore, the difference among the three dust models allows us to estimate the systematic uncertainties in the observed SFRs.
\par
Figure \ref{plot_CSFD} illustrates the predicted cosmic star formation rate density (SFRD) for the best-fit models of M-SFR, M-DTG and M-GCD. Their values are listed in Table \ref{tab_CSFD}. We compare the results with \cite{2015ApJ...803...34B}, whose estimations are based on the CMRs of \cite{2014ApJ...793..115B} and the \cite{1999ApJ...521...64M} relation. Since the results of \cite{2015ApJ...803...34B} and our models use the same observational information, the comparison between them quantifies the systematic errors of using the \cite{1999ApJ...521...64M} relation with respect to correcting the dust extinction. We note that all our models suggest bluer intrinsic UV continuum slopes than the one used in \cite{1999ApJ...521...64M} as shown in Figure \ref{best-fit}. Figure \ref{plot_IRX} also illustrates that the best-fit results of M-DTG and M-GCD have shallower IRX - $\beta$ relations. Thus, compared with the \citet{1999ApJ...521...64M} relation, the dust extinction in these two models is stronger for bluer galaxies but weaker for redder galaxies. On the other hand, the dust attenuation is stronger for all galaxies in the best-fit M-SFR. It can be seen from Figure \ref{plot_CSFD} that the cosmic SFRD of the best-fit M-DTG and M-GCD are consistent with those of \cite{2015ApJ...803...34B}, while the results of the best-fit M-SFR is roughly a factor of two higher. We also compare our results with \cite{2018MNRAS.475.2891D}. Their dust-corrected SFRs are derived from the energy balance SED-fitting code \textsc{magphys} \citep{2008MNRAS.388.1595D}. Better consistency is found between their measurements and our best-fit models of M-DTG and M-GCD, given the size of the error bars on those points. Overall, Figure \ref{plot_CSFD} suggests that uncertainty in the dust relations introduces at least a factor of two systematic error into the inferred cosmic SFRD at $z \gtrsim 6$.  

\begin{table}
    \resizebox{\columnwidth}{!}{%
    \begin{tabular}{ccccccc}
    \hline \hline
    z & \multicolumn{3}{c}{All galaxies} & \multicolumn{3}{c}{$M_\text{UV} < -17$} \\
      & M-SFR       & M-DTG       & M-GCD       & M-SFR     & M-DTG     & M-GCD    \\ \hline
    4 & -1.04       & -1.24       & -1.28       & -1.07     & -1.29     & -1.32    \\
    5 & -1.19       & -1.44       & -1.44       & -1.25     & -1.51     & -1.52    \\
    6 & -1.38       & -1.62       & -1.64       & -1.46     & -1.72     & -1.72    \\
    7 & -1.56       & -1.78       & -1.83       & -1.73     & -1.93     & -1.97    \\ \hline
    \end{tabular}
    }%
    \caption{Tabular data of predicted cosmic star formation rate density (SFRD) for the three different dust models. These values are plotted in Figure \ref{CSFD} and are in a unit of $\log_{10} ( \rho_\text{SFR}/ (\solarmass \text{yr}^{-1} \text{Mpc}^{-3}))$.}
    \label{tab_CSFD}
\end{table}

\section{Summary} \label{sum}
This work investigates the IRX - $\beta$ relation and cosmic SFRD at $z \sim 4 - 7$ by combining the the \textsc{meraxes} semi-analytic galaxy formation model \citep{2016MNRAS.462..250M,2017MNRAS.472.2009Q} and the \cite{2000ApJ...539..718C} dust attenuation model. The supernova feedback model of \textsc{meraxes} is updated using results from previous studies \citep{2015MNRAS.454.2691M,2016MNRAS.461.1760H,2018MNRAS.479....2C}, which aim to reproduce the redshift evolution of the mass metallicity relation. We introduce three different parametrisations of the dust optical depths, which are related to the star formation rate (M-SFR), dust-to-gas ratio (M-DTG) and gas column density (M-GCD) respectively. These lead to five free parameters in each dust model in additional to those in \textsc{meraxes}.
\par
The determinations on not only the dust parameters but also the \textsc{meraxes} free parameters constitute the primary part of this work. For galaxy formation parameters, we focus on the star formation efficiency, critical mass, mass loading factor and supernova coupling efficiency. We adopt a Bayesian approach, calibrating these parameters against the UV luminosity functions (LFs) of \cite{2015ApJ...803...34B} and colour magnitude relations (CMRs) of \cite{2014ApJ...793..115B} at $z \sim 4 - 7$. The posterior distribution of these parameters is estimated using multimodal nested sampling \citep{2009MNRAS.398.1601F}. We find that these observations can be fit extremely well by all the three dust models. However, the preferred parameter ranges are quite different among the three dust models. Our analysis indicates that the combination of the LFs and CMRs can put strong constraints on a given dust attenuation model, since the model is required to reproduce the shape of both observations. The differences in our results are due to the different assumptions of the dust models, which results in different relations between UV dust attenuation and intrinsic UV magnitude.
\par
We then demonstrate the predictions of our calibration results. Using energy balance arguments, we estimate the IRX for each model galaxy. We find that the predicted IRX - $\beta$ relations are quite different from the \cite{1999ApJ...521...64M} relation, and contain large intrinsic scatter, which might explain the current discrepancy among several high redshift observations \citep[e.g.][]{2015Natur.522..455C,2016A&A...587A.122A,2016ApJ...833...72B,2017ApJ...845...41B,2017MNRAS.472..483F,2018MNRAS.479.4355K}. We also confirm the correlation between the intrinsic scatter and sSFR. This finding is consistent with other theoretical studies \citep{2017MNRAS.472.2315P,2017ApJ...840...15S,2018MNRAS.474.1718N,2019arXiv190101747C}. Secondly, we present model predictions for the cosmic SFRD, and compare these with the observations of \cite{2015ApJ...803...34B} and \cite{2018MNRAS.475.2891D}. The difference among the three dust models implies at least a factor of two systematic uncertainty in the observed SFRD when corrected using the Meurer IRX - $\beta$ relation.
\par
This work has simultaneously constrained the free parameters of a semi-analytic galaxy formation model and additional dust parameters using observations of UV properties. Within a Bayesian framework, our approach establishes a more direct connection between the model and observations despite the complexity. This approach is particularly useful for studies at high redshifts where UV properties are the most robust observables. This work could be further improved by explicitly modelling the dust evolution \citep[e.g][]{2016MNRAS.462.3130M,2017MNRAS.471.3152P,2018PhR...780....1D}, which might reduce the systematic uncertainties due to different assumptions in the dust models. Additional free parameters (e.g. the time scale of dust growth) in such dust evolution models could also be constrained using our methodology.

\section*{acknowledgement}
This research was supported by the Australian Research Council Centre of Excellence for All Sky Astrophysics in 3 Dimensions (ASTRO 3D), through project	number CE170100013. This work was performed on the OzSTAR national facility at Swinburne University of Technology. OzSTAR is funded by Swinburne University of Technology and the National Collaborative Research Infrastructure Strategy (NCRIS). We thank the referee for providing constructive comments to improve the quailty of this paper. 
\par
In addition to those already been mentioned in the paper, we also acknowledge the use of the following software: \textsc{astropy} \footnote{http://www.astropy.org} \citep{astropy:2013, astropy:2018}, \textsc{corner} \citep{2016JOSS....1...24F}, \textsc{cython} \citep{2011CSE....13b..31B}, \textsc{ipython} \citep{2007CSE.....9c..21P}, \textsc{matplotlib} \citep{2007CSE.....9...90H}, \textsc{numpy} \citep{2011CSE....13b..22V}, \textsc{pandas} \citep{mckinney-proc-scipy-2010}, \textsc{seaborn}\footnote{https://github.com/mwaskom/seaborn} and \textsc{scipy} \citep{scipy}.

\bibliographystyle{mnras}
\bibliography{reference}

\appendix
\section{Posterior distributions}
\begin{figure*}
 \includegraphics[trim=0 -5 0 -5cm,clip,width=\textwidth]{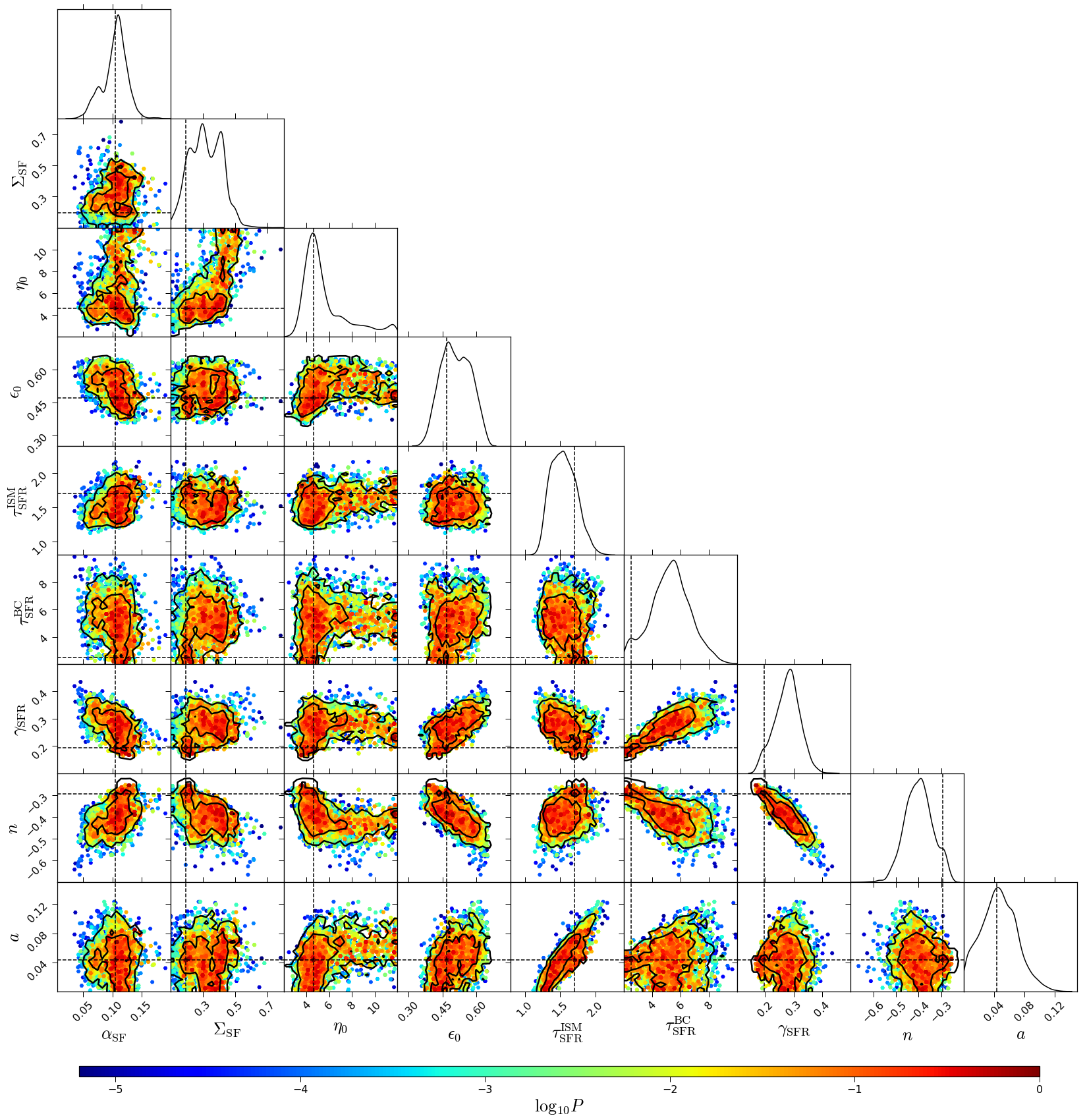}
 \caption{Posterior distribution of the galaxy and dust parameters for \textsc{meraxes} with a star formation rate dependent (SFR) dust model. We refer the model to as M-SFR, which is described in Section \ref{SFR_model}. The posterior distribution is a function of star formation efficiency $\alpha_\text{SF}$, critical mass normalisation $\Sigma_\text{SF}$, mass loading factor $\eta_0$, supernova energy coupling efficiency $\epsilon_0$, optical depth normalisations of interstellar media $\tau^\text{ISM}_\text{SFR}$ and birth cloud $\tau^\text{BC}_\text{SFR}$, optical depth scaling of star formation rate $\gamma_\text{SFR}$, reddening slope $n$ and optical depth redshift dependence $a$. See also Table \ref{params} for a summary of these parameters. Diagonal panels show the one parameter marginalised distributions. In the off-diagonal panels, solid black lines are the 68\% and 95\% contours of the two parameter marginalised distributions. Colour points reflect the values of the posterior distribution, and the maximum is normalised to unity. The point that has the highest value is chosen to be best-fit results, which is specified by the dashed lines. Their values are listed in Table \ref{params}.}
 \label{pdf_SFR}
\end{figure*}

\begin{figure*}
 \includegraphics[trim=0 -5 0 -5cm,clip,width=\textwidth]{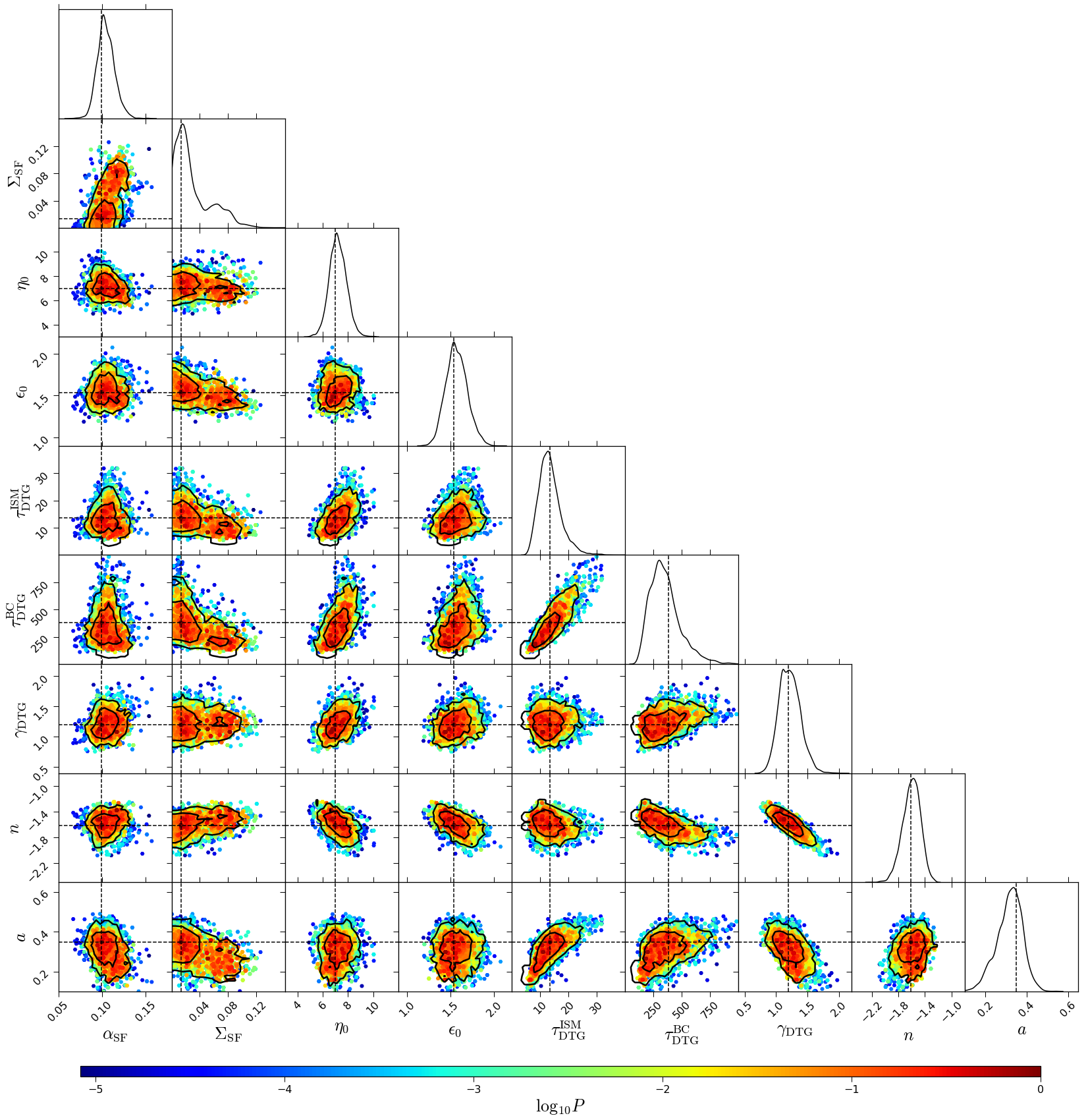}
 \caption{Posterior distribution of the galaxy and dust parameters for \textsc{meraxes} with a dust-to-gas ratio (DTG) dependent dust model. The model is referred to as M-DTG and described in Section \ref{DTG_model}. The posterior distribution is a function of star formation efficiency $\alpha_\text{SF}$, critical mass normalisation $\Sigma_\text{SF}$, mass loading factor $\eta_0$, supernova energy coupling efficiency $\epsilon_0$, optical depth normalisations of interstellar media $\tau^\text{ISM}_\text{DTG}$ and birth cloud $\tau^\text{BC}_\text{DTG}$, slope of the dust-to-gas ratio $\gamma_\text{DTG}$, reddening slope $n$ and optical depth redshift dependence $a$. See also Table \ref{params} for a summary of these parameters. Diagonal panels show the one parameter marginalised distributions. In the off-diagonal panels, solid black lines are the 68\% and 95\% contours of the two parameter marginalised distributions. Colour points reflect the values of the posterior distribution, and the maximum is normalised to unity. The point that has the highest value is chosen to be best-fit results, which is specified by the dashed lines. Their values are listed in Table \ref{params}.}
 \label{pdf_DTG}
\end{figure*}

\begin{figure*}
 \includegraphics[trim=0 -5 0 -5cm,clip,width=\textwidth]{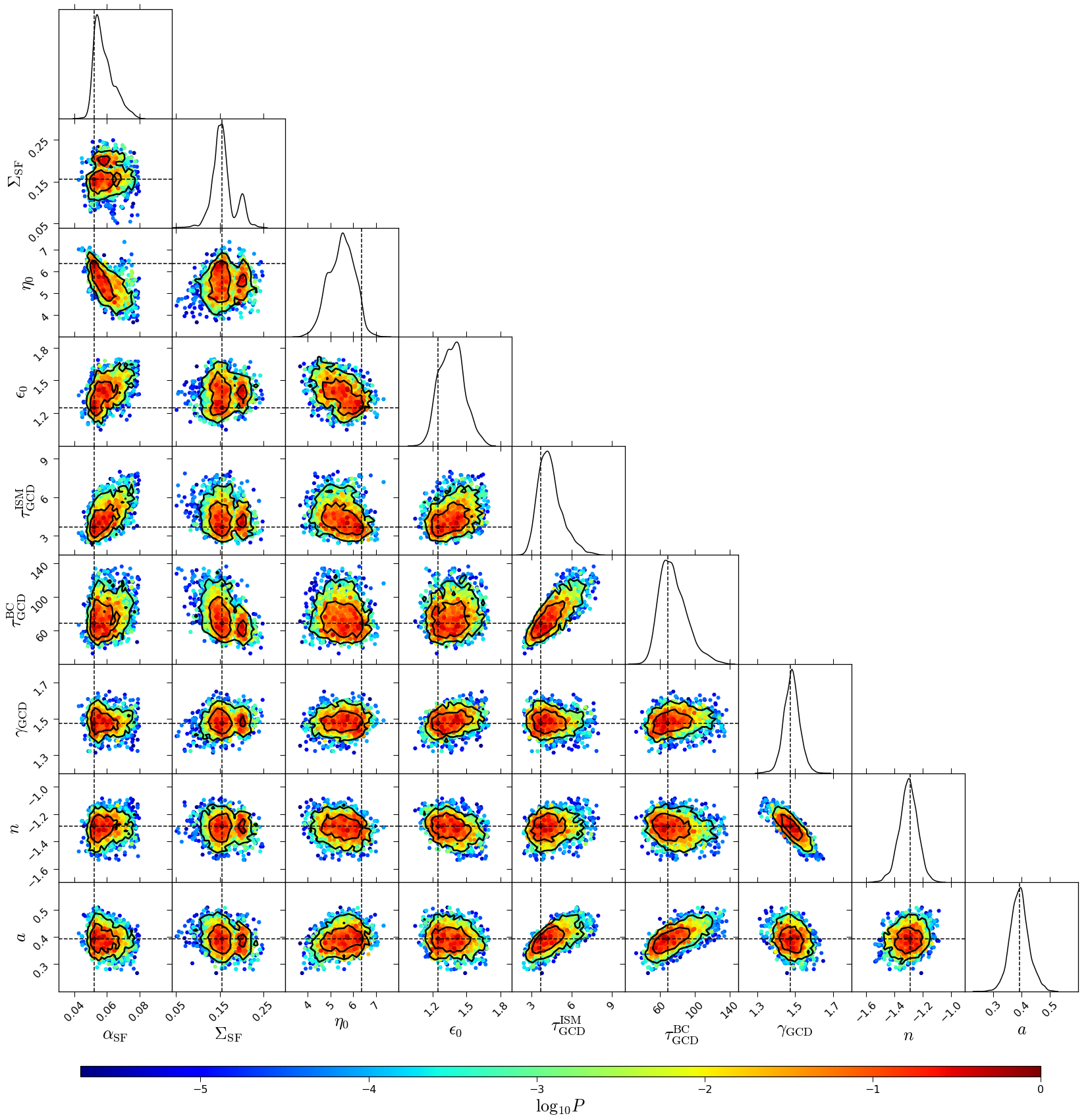}
 \caption{Posterior distribution of the galaxy and dust parameters for \textsc{meraxes} with a gas column density (GCD) dependent dust model. The model is referred to as M-GCD and described in Section \ref{GCD_model}. The posterior distribution is a function of star formation efficiency $\alpha_\text{SF}$, critical mass normalisation $\Sigma_\text{SF}$, mass loading factor $\eta_0$, supernova energy coupling efficiency $\epsilon_0$, optical depth normalisations of interstellar media $\tau^\text{ISM}_\text{GCD}$ and birth cloud $\tau^\text{BC}_\text{GCD}$, optical depth scaling of gas mass $\gamma_\text{GCD}$, reddening slope $n$ and optical depth redshift dependence $a$. See also Table \ref{params} for a summary of these parameters. Diagonal panels show the one parameter marginalised distributions. In the off-diagonal panels, solid black lines are the 68\% and 95\% contours of the two parameter marginalised distributions. Colour points reflect the values of the posterior distribution, and the maximum is normalised to unity. The point that has the highest value is chosen to be best-fit results, which is specified by the dashed lines. Their values are listed in Table \ref{params}.}
 \label{pdf_GCD}
\end{figure*}

This appendix illustrates the posterior distributions of M-SFR, M-DTG and M-GCD in Figures \ref{pdf_SFR}, \ref{pdf_DTG} and \ref{pdf_GCD}, respectively. These results are obtained using the methodology introduced in Section \ref{calib}.

\bsp
\label{lastpage}
\end{document}